\pdfoutput=1

\documentclass[12pt]{article}%
\usepackage{amsmath}
\usepackage{amsfonts}
\usepackage{amssymb}
\usepackage{graphicx}%
\providecommand{\U}[1]{\protect\rule{.1in}{.1in}}
\begin{document}
\begin{titlepage}
\vspace{.3cm} \vspace{1cm}
\begin{center}
\baselineskip=16pt \centerline{\Large\bf Quanta of Geometry and Unification} \vspace{2truecm} \centerline{\large\bf Ali H.
Chamseddine$^{1,2}$ } \vspace{.5truecm}
\emph{\centerline{$^{1}$Physics Department, American University of Beirut, Lebanon}}
\emph{\centerline{$^{2}$I.H.E.S. F-91440 Bures-sur-Yvette, France}}
\end{center}
\vspace{2cm}
\begin{center}
{\bf Abstract}
\end{center}
This is a tribute to Abdus Salam's memory whose insight and creative thinking
set for me a role model to follow. In this contribution I show that the simple
requirement of volume quantization in space-time (with Euclidean signature)
uniquely determines the geometry to be that of a noncommutative space whose
finite part is based on an algebra that leads to Pati-Salam grand unified
models. The Standard Model corresponds to a special case where a mathematical
constraint (order one condition) is satisfied. This provides evidence that
Salam was a visionary who was generations ahead of his time.

\end{titlepage}

\section{Introduction}

In 1973 I got a scholarship from government of Lebanon to pursue my graduate
studies at Imperial College, London. Shortly after I\ arrived, I was walking
through the corridor of the theoretical physics group, I saw the name Abdus
Salam on a door. At that time my information about research in theoretical
physics was zero, and since Salam is an Arabic name, and the prime minister in
Lebanon at that time was also called Salam, I\ knocked at his door and asked
him whether he is Lebanese. He laughed and explained to me that he is from
Pakistan. He then asked me why I\ wanted to study theoretical physics. I said
the reason is that I love mathematics. He smiled and told me that I am in the
wrong department. In June 1974, having finished the Diploma exams I\ asked
Salam to be my Ph.D. advisor and he immediately accepted and gave me two
preprints to read and to chose one of them as my research topic. The first
paper was with Strathdee \cite{Strathdee} on the newly established field of
supersymmetry (a word he coined), and the other is his paper with Pati
\cite{Pati} on the first Grand Unification model, now known as the Pati-Salam
model. Few days later I came back and told Salam that I have chosen
supersymmetry which I thought to be new and promising. Little I knew that the
second project will come back to me forty years later from studying the
geometric structure of space-time, as will be explained in what follows. In
this respect, Salam was blessed with amazing foresight. I realized this early
enough. In 1974 I got him a metallic Plaque with Arabic Calligraphy engraving
from Koran which translates "We opened for thee a manifest victory".%
\begin{figure}[ptb]%
\centering
\includegraphics[
natheight=2.187100in,
natwidth=2.500200in,
height=2.2278in,
width=2.5417in
]%
{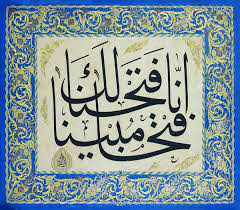}%
\caption{We opened for thee a manifest victory}%
\end{figure}
Indeed soon after the \ year 1974, the Salam's path to victory and great fame
has already started. His influence on me, although brief, was very strong. He
was to me the role model of a dedicated scientist pioneering in helping fellow
scientists, especially those who come from developing countries. For his
insight and kindness I am eternally grateful.

\section{Volume Quantization}

The work I\ am reporting in this presentation is the result of a long-term
collaboration with Alain Connes over a span of twenty years starting in 1996
\cite{AC2} \cite{saddam} \cite{CC} \cite{CC2} \cite{AC} \cite{Reselience}. In
the latest work on volume quantization we were joined by Slava Mukhanov
\cite{CCM} \cite{CCM2}. On the inner fluctuations of the Dirac operator over
automorphisms of the noncommutative algebra times its opposite, we were joined
by Walter von Suijlekom \cite{CCS} \cite{CCS2} \cite{CCS3}.

At very small distances, of the order of Planck length $1.6\times10^{-35}$m,
we expect the nature of space-time to change. It is then natural to ask
whether there is a fundamental unit of volume in terms of which the volume of
space is quantized. The present volume of space is of the order of $10^{60}$
Planck units \cite{CC}, but at the time of the big bang, this volume must have
been much smaller. To explore this idea, we start with the observation that it
is always possible to define a map $Y^{A}$, $A=1,\cdots,n+1,$ from an $n$
dimensional manifold $M_{n}$ to the $n-$sphere $S^{n}$ so that $Y^{A}Y^{A}=1$
\cite{Greub}. This map has a degree, the winding number over the sphere, which
is an integer. This is defined in terms of an $n$-form $\omega_{n}$, the
integral of which is a topological invariant%
\begin{equation}
\omega_{n}=\frac{1}{n!}\epsilon_{A_{1}A_{2}\cdots A_{n+1}}Y^{A_{1}}dY^{A_{2}%
}\cdots dY^{A_{n+1}}%
\end{equation}
If we equate $\omega_{n}$ with the volume form
\begin{equation}
v_{n}=\sqrt{g}dx^{1}\wedge dx^{2}\cdots\wedge dx^{n}%
\end{equation}
so that
\begin{equation}%
{\displaystyle\int\limits_{M_{n}}}
v_{n}=%
{\displaystyle\int\limits_{M_{n}}}
w_{n}=\deg\left(  Y\right)  \in\mathbb{Z} \label{winding}%
\end{equation}
then the volume of the manifold $M_{n}$ will be quantized \cite{Greub} and
given by an integer multiple of the unit sphere in Planck units. This
hypothesis, however, has topological obstruction. For the equality to hold,
the pullback of the volume form is a four-form that does not vanish anywhere,
and thus the Jacobian of the map $Y^{A}$ does not vanish anywhere and is then
a covering of the sphere $S^{n}.$ However, since the sphere is simply
connected, the manifold $M_{n}$ must be a covering of the sphere $S^{n}$
\cite{CMV} \cite{CCM2}. This implies that the manifold must be disconnected,
each component being a sphere. This gives a bubble picture of space, where
every bubble of Planckian volume has the topology of a sphere $S^{n}.$ This is
not an attractive picture because one would have to invent a mechanism for
condensation of the bubbles at lower energies. To rescue this idea, we first
rewrite the proposal (\ref{winding}) in a different form. Let $D_{0}$ be the
Dirac operator on $M_{n}$ given by
\begin{equation}
D_{0}=\gamma^{a}e_{a}^{\mu}\left(  \partial_{\mu}+\frac{1}{4}\omega_{\mu}%
^{bc}\gamma_{bc}\right)  \label{Dzero}%
\end{equation}
where $e_{a}^{\mu}$ is the (inverse) vielbein and $\omega_{\mu}^{bc}$ is the
spin-connection. Notice that in momentum space, the Dirac operator could be
identified with momenta $p_{a}$, Feynman slashed with the Clifford algebra
spanned by $\gamma^{a}.$ In analogy, introduce a new Clifford algebra
$\Gamma^{A}$ such that
\begin{equation}
\left\{  \Gamma_{A},\Gamma_{B}\right\}  =2\kappa\delta_{AB},\qquad\kappa
=\pm1,\qquad A=1,\cdots,n+1 \label{Cliff}%
\end{equation}
and slash the coordinates $Y^{A}$ with $\Gamma_{A}$%
\begin{equation}
Y=Y^{A}\Gamma_{A},\qquad Y=Y^{\ast},\qquad Y^{2}=1
\end{equation}
then a compact way of writing equation (\ref{winding}) is given by%
\begin{equation}
\left\langle Y\left[  D_{0},Y\right]  ^{n}\right\rangle =\gamma,\qquad
n=\mathrm{even} \label{Heis}%
\end{equation}
where $\gamma$ is the chirality operator on $M_{n},$ $\gamma=\gamma_{1}%
\cdots\gamma_{n}$ and $\left\langle {}\right\rangle $ denotes taking the trace
over the Clifford algebra spanned by $\Gamma^{A}.$ In this form, the
quantization condition is a generalization of the Heisenberg commutator for
momenta and coordinates $\left[  p,x\right]  =-i\hbar.$ It is also identical
to the Chern character formula in noncommutative geometry, which is a special
case of the orientability condition with idempotent elements. This suggests to
consider the above proposal for a noncommutative space defined by a spectral
triple $\left(  \mathcal{A},\mathcal{H},D\right)  $ together with reality
operator $J$ and chirality $\gamma$ \cite{Connesbook} \cite{Connes}
\cite{CMos}. Here $\mathcal{A}$ is an associative $\ast$ algebra with
involution and unit element, $\mathcal{H}$ a Hilbert space, $D$ is a
self-adjoint operator with bounded spectrum for $\left(  D^{2}+1\right)
^{-1}.$ The chirality operator commutes with the algebra $\mathcal{A},$
$\gamma a=a\gamma,$ $\forall a\in\mathcal{A}.$ The following properties are
assumed to hold%
\begin{equation}
J^{2}=\epsilon,\qquad JD=\epsilon^{\prime}DJ,\qquad J\gamma=\epsilon"\gamma
J,\qquad\epsilon,\epsilon^{\prime},\epsilon"\in\left\{  -1,1\right\}
\end{equation}
which defines a $KO$ dimension (mod $8$) of the noncommutative space. As an
example, for a Riemannian manifold $\mathcal{A}=C^{\infty}\left(  M\right)  ,$
$\mathcal{H=}L^{2}\left(  S\right)  ,$ $D$ is the Dirac operator
(\ref{Dzero}), $\gamma$ is the chirality, $J$ is the charge conjugation operator.

The operator $J$ sends the algebra $\mathcal{A}$ into its commutant
$\mathcal{A}^{o}$
\begin{equation}
\left[  a,Jb^{\ast}J^{-1}\right]  =0,\qquad\forall a,b\in\mathcal{A}%
\end{equation}
where $Ja^{\ast}J^{-1}\in\mathcal{A}^{o}$, and thus the left action and the
right action acting on elements of the Hilbert space commute.

Going back to the volume quantization condition, the slashed coordinates $Y$
when acted on with the $J$ become $Y^{\prime}=JYJ^{-1}$ which commutes with it
$\left[  Y,Y^{\prime}\right]  =0.$ If $Y$ is slashed with the Clifford algebra
(\ref{Cliff}) where $\kappa=1$ so that $Y^{\prime}$ will correspond to the
Clifford algebra with $\kappa=-1.$ It is essential to have a volume
quantization condition involving both $Y$ and $Y^{\prime}.$ To do this, let
$e=\frac{1}{2}\left(  Y+1\right)  $ so that $e^{2}=e$ and similarly
$e^{\prime}=\frac{1}{2}\left(  Y^{\prime}+1\right)  $ with $e^{\prime
2}=e^{\prime}.$ The \ product $E=ee^{\prime}$ also satisfies $E^{2}=E$ which
implies that the composite coordinate $Z=2E-1$ satisfies $Z^{2}=1.$ This
suggests that we modify our volume quantization condition (\ref{Heis}) to
become \cite{CCM}%
\begin{equation}
\left\langle Z\left[  D_{0},Z\right]  ^{n}\right\rangle =\gamma,\qquad
n=\mathrm{even} \label{quanta}%
\end{equation}
For dimensions $n=2$ and $n=4$ this relation splits into two pieces, one is a
function of $Y$ and the other a function of $Y^{\prime}$ \cite{CCM2}
\begin{equation}
\left\langle Y\left[  D_{0},Y\right]  ^{n}\right\rangle +\left\langle
Y^{\prime}\left[  D_{0},Y^{\prime}\right]  ^{n}\right\rangle =\gamma,\qquad
n=2,4 \label{windingsum}%
\end{equation}
For $n=6$ there are mixing terms between $Y$ and $Y^{\prime}$ and the relation
(\ref{quanta}) does not factorize. Thus the only realistic case where we take
the volume quantization condition (\ref{quanta}) to hold is for manifolds of
dimension $n=4.$ In what follows we restrict our considerations to dimensions
$n=4$, where we will find out the special importance of the number $4$.

\section{Noncommutative space}

From now on we specialize to dimension $n=4.$ In this case, (\ref{windingsum})
holds and with this condition, we prove that if $M_{4}$ is an oriented
four-manifold then a solution of the two sided equation (\ref{quanta}) exists
and is equivalent to the existence of the two maps $Y$ and $Y^{\prime}%
:M_{4}\rightarrow S^{4}$ such that the sum of the two pullbacks $Y^{\ast
}\left(  \omega\right)  +Y^{^{\prime}\ast}\left(  \omega\right)  $ does not
vanish anywhere and
\begin{equation}
\mathrm{vol}\left(  M\right)  =\left(  \left(  \deg Y\right)  +\left(  \deg
Y^{\prime}\right)  \right)  \mathrm{vol}\left(  S^{4}\right)
\end{equation}
The proof is difficult because in four dimensions, the kernel of the map $Y$
is of codimension $2.$ Details are given in reference \cite{CCM2}.
Fortunately, for this relation to hold, the only conditions on the manifold
$M_{4}$ is the vanishing of the second Steifel-Whitney class $w_{2},$ which is
automatically satisfied for spin manifolds, and that $\mathrm{vol}\left(
M\right)  $ should be larger than four units \cite{CCM2}. In this setting, the
four dimensional manifold emerges as a composite of the inverse maps of the
product of two spheres of Planck size. The manifold $M_{4}$ which is folded
many times in the product, unfolds to macroscopic size. The two different
spheres, associated with the two Clifford algebras can be considered as quanta
of geometry which are the building blocks to generate an arbitrary oriented
four dimensional spin-manifold. We can show that the manifold $M_{4}$, the two
spheres with their maps $Y,$ $Y^{\prime}$ and their associated Clifford
algebras define a noncommutative space which is the basis of unification of
all fundamental interactions, including gravity.

To study this noncommutative space, we first note that the Clifford algebras
with $\kappa=1$ and $\kappa=-1$ are given by \cite{LM}
\begin{align}
\mathrm{Cliff}\left(  +,+,+,+,+\right)   &  =M_{2}\left(  \mathbb{H}\right) \\
\mathrm{Cliff}\left(  -,-,-,-,-\right)   &  =M_{4}\left(  \mathbb{C}\right)
\end{align}
and thus the algebra $\mathcal{A}_{F}$ of the finite space is
\begin{equation}
\mathcal{A}_{F}=M_{2}\left(  \mathbb{H}\right)  \oplus M_{4}\left(
\mathbb{C}\right)  \label{finite}%
\end{equation}
The finite Hilbert space $\mathcal{H}_{F}$ is then the basic representation
$\left(  4,4\right)  $ where the the first $4$ is acted on by the matrix
elements $M_{2}\left(  \mathbb{H}\right)  $ and the second $4$ is acted on by
the matrix elements $M_{4}\left(  \mathbb{C}\right)  .$ It is tantalizing to
observe that the same finite algebra (\ref{finite}) was obtained by
classifying all finite algebras of $KO$ dimension $6$, required to avoid
mirror fermions. The maps $Y$ and $Y^{\prime}$ are functions of the
coordinates $x^{\mu}.$ Since $Y^{2}=1$ composing words from the elements of
the algebra $M_{2}\left(  \mathbb{H}\right)  $ and $Y$ of the form
$a_{1}Ya_{2}Y\cdots a_{i}Y,$ $\forall i,$ and similarly for $Y^{\prime}$ will
generate the algebra
\begin{align}
\mathcal{A}  &  =C^{\infty}\left(  M_{4},M_{2}\left(  \mathbb{H}\right)
\oplus M_{4}\left(  \mathbb{C}\right)  \right) \\
&  =C^{\infty}\left(  M_{4},\right)  \otimes\left(  M_{2}\left(
\mathbb{H}\right)  \oplus M_{4}\left(  \mathbb{C}\right)  \right)
\end{align}
The associated Hilbert space is then%
\begin{equation}
\mathcal{H}=L^{2}\left(  S\right)  \otimes\mathcal{H}_{F}%
\end{equation}
and the Dirac operator is
\begin{equation}
D=D_{0}\otimes1+\gamma_{5}\otimes D_{F}%
\end{equation}
where $D_{F}$ is a self-adjoint matrix operator acting on $\mathcal{H}_{F}.$
The reality operator $J_{F}$ acting on the finite algebra $\mathcal{A}_{F}$
satisfies
\begin{equation}
J_{F}\left(  x,y\right)  =\left(  y^{\ast},x^{\ast}\right)
\end{equation}
so that the reality operator $J$ will be
\begin{equation}
J=C\otimes J_{F}%
\end{equation}
where $C$ is the charge conjugation operator. Finally, the chirality operator
$\gamma$ is given by%
\begin{equation}
\gamma=\gamma_{5}\otimes\gamma_{F}%
\end{equation}
One then finds that the finite noncommutative space $\left(  \mathcal{A}%
_{F},\mathcal{H}_{F},D_{F}\right)  $ with $J_{F}$ and $\gamma_{F}$ have a $KO$
dimension $6.$ Thus the $KO$ dimension of the full noncommutative space is
$10$ \cite{Co} \cite{Barrett}$.$ Elements of the Hilbert space $\mathcal{H}$
are of the form
\begin{equation}
\Psi=\left(
\begin{array}
[c]{c}%
\psi\\
\psi^{c}%
\end{array}
\right)
\end{equation}
which can be denoted by $\Psi_{\widehat{\alpha}\alpha I}$ where
$\widehat{\alpha}=1,\cdots,4$ transforms as a space-time spinor,
$\alpha=1,\cdots,4$ transforms as a $4$ under the action of $M_{2}\left(
\mathbb{H}\right)  $ and $I=1,\cdots,4$ transforms as a $4$ under the action
of $M_{4}\left(  \mathbb{C}\right)  .$ Thus $\Psi$ represents a $16$
space-time Dirac spinor and its conjugate. \ However, since the $KO$ dimension
of the full space is $10,$ this allows to impose the following two conditions
(in the Lorentzian version) on $\Psi$, the chirality and the reality
conditions%
\begin{equation}
\gamma\Psi=\Psi,\qquad J\Psi=\Psi
\end{equation}
which implies that $\psi$ is a chiral $16$ and $\psi^{c}$ is not an
independent spinor, and is given by $C\overline{\psi}^{T}$. \ Now, the
$\gamma_{F}$ chirality operator must commute with $\mathcal{A}_{F}.$ If this
operator is taken to act on the first algebra $M_{2}\left(  \mathbb{H}\right)
$, this implies that only the subalgebra $\mathbb{H}_{R}\oplus\mathbb{H}_{L}$
is preserved. The $16$ spinor then transforms under the finite algebra
\begin{equation}
\mathcal{A}_{F}=\left(  \mathbb{H}_{R}\oplus\mathbb{H}_{L}\right)  \oplus
M_{4}\left(  \mathbb{C}\right)  \label{PS}%
\end{equation}
as $\left(  2_{R},1_{L},4\right)  +\left(  1_{R},2_{L},4\right)  .$ Elements
of the Hilbert space do transform under automorphisms of the algebra
$\mathcal{A}\otimes\mathcal{A}^{o}$ as%
\begin{equation}
\Psi\rightarrow U\Psi,\qquad U=u\,\widehat{u},\qquad u\in\mathcal{A}%
,\qquad\widehat{u}=Ju^{\ast}J^{-1}\in\mathcal{A}^{o}%
\end{equation}
The action of the Dirac operator $D$ on elements of the Hilbert space $\Psi$
does not transform covariantly under the automorphisms $U.$ It can be made so
by adding to $D$ a connection $A$ so that
\begin{equation}
D_{A}=D+A \label{Dirac}%
\end{equation}
such that
\begin{equation}
D_{A}\left(  U\Psi\right)  =UD_{A^{u}}\Psi
\end{equation}
This fixes the connection $A$ to be given by%
\begin{equation}
A=%
{\displaystyle\sum}
a\widehat{a}\left[  D,b\widehat{b}\right]
\end{equation}
which can be decomposed into three parts
\begin{equation}
A=A^{\left(  1\right)  }+JA^{\left(  1\right)  }J^{-1}+A^{\left(  2\right)  }
\label{connection}%
\end{equation}
where
\begin{align}
A^{\left(  1\right)  }  &  =%
{\displaystyle\sum}
a\left[  D,b\right] \\
A^{\left(  2\right)  }  &  =%
{\displaystyle\sum}
\widehat{a}\left[  A^{\left(  1\right)  },\widehat{b}\right]
\end{align}
The connection $A$ transforms as
\begin{align}
A^{\left(  1\right)  u}  &  =uA^{\left(  1\right)  }u^{\ast}+u\left[
D,u^{\ast}\right] \\
A^{\left(  2\right)  u}  &  =\widehat{u}A^{\left(  1\right)  }\widehat{u}%
^{\ast}+\widehat{u}\left[  u\left[  D,u^{\ast}\right]  ,\widehat{u}^{\ast
}\right]
\end{align}
In the special case where the order one condition is satisfied,%
\begin{equation}
\left[  a,\left[  D,\widehat{b}\right]  \right]  =0,\qquad\forall
a,b\in\mathcal{A} \label{order}%
\end{equation}
this implies that
\begin{equation}
A^{\left(  2\right)  }=0
\end{equation}
The condition (\ref{order}) restricts the algebra $\left(  \mathbb{H}%
_{R}\oplus\mathbb{H}_{L}\right)  \oplus M_{4}\left(  \mathbb{C}\right)  $ to
the subalgebra
\begin{equation}
\mathcal{A}_{F}=\mathbb{C}\oplus\mathbb{H}_{L}\oplus M_{3}\left(
\mathbb{C}\right)  \label{SM}%
\end{equation}
where the algebra $\mathbb{C}$ is embedded in the diagonal part of
$\mathbb{H}_{R}\oplus M_{4}\left(  \mathbb{C}\right)  .$ This is the algebra
that gives rise to the Standard Model.

In practical terms, the connection (\ref{connection}) can be calculated using
simple matrix algebra. The results show that the elements of the connection
along the three separate algebras $\mathbb{H}_{R}\oplus\mathbb{H}_{L}\oplus
M_{4}\left(  \mathbb{C}\right)  $ are tensored with the space-time
$\gamma^{\mu}$ and are the gauge fields of $SU\left(  2\right)  _{R}$,
$SU\left(  2\right)  _{L},$ and $SU\left(  4\right)  $ which are those of the
Pati-Salam models. Components of the connection along the off-diagonal
elements between the three different algebras are tensored with the space-time
chirality $\gamma_{5}$ and are the Higgs fields. Representations of the Higgs
fields depend on the form of the initial finite space Dirac operator and fall
into three different classes of Pati-Salam models. The first class have the
Higgs fields in the $SU\left(  2\right)  _{R}\times SU\left(  2\right)
_{L}\times SU\left(  4\right)  $ representations
\begin{align}
\Sigma_{aI}^{\overset{.}{b}J}  &  =\left(  2,2,1\right)  +\left(
1,1,1+15\right)  ,\\
H_{aIbJ}  &  =\left(  1,1,6\right)  +\left(  1,3,10\right)  ,\qquad\\
H_{\overset{.}{a}I\overset{.}{b}J}  &  =\left(  1,1,6\right)  +\left(
3,1,10\right)
\end{align}
For the second class we have the same fields as the first class with the
restriction $H_{aIbJ}=0.$ The third class is a special case of the second
class, but where the fields $\Sigma_{aI}^{\overset{.}{b}J}$ and
$H_{\overset{.}{a}I\overset{.}{b}J}$ are composites of more fundamental
fields
\begin{equation}
H_{\overset{.}{a}I\overset{.}{b}J}=\Delta_{\overset{.}{a}J}\Delta
_{\overset{.}{b}I},\qquad\Sigma_{aI}^{\overset{.}{b}J}\sim\phi_{a}%
^{\overset{.}{b}}\Sigma_{I}^{J}%
\end{equation}
where $\Delta_{\overset{.}{a}J}$ is in the $\left(  2,1,4\right)  $
representation, $\phi_{a}^{\overset{.}{b}}$ is in the $\left(  2,2,1\right)  $
representation and $\Sigma_{I}^{J}$ is in the $\left(  1,1,1+15\right)  .$

When the order one condition (\ref{order}) on the algebra is satisfied, the
algebra of the finite noncommutative space reduces to the subalgebra
(\ref{SM}). Components of the connection along the separate three algebras are
tensored with $\gamma^{\mu}$ and are those of $U\left(  1\right)  _{Y}\times
SU\left(  2\right)  \times SU\left(  3\right)  .$ The component of the
connection along the off-diagonal elements of the algebras $\mathbb{C}$ and
$\mathbb{H}$ tensored with $\gamma_{5}$ is the Higgs doublet. There is also a
singlet component, tensored with $\gamma_{5}$ and connects the right-handed
neutrino to its conjugate. The $16$ fermions have the representations under
$SU\left(  2\right)  _{R}\times SU\left(  2\right)  _{L}\times SU\left(
4\right)  $ given by%
\begin{equation}
\left(  2,1,4\right)  +\left(  1,2,4\right)
\end{equation}
In the special case of the subalgebra is that of the SM, the fermion
representation with respect to $U\left(  1\right)  _{Y}\times SU\left(
2\right)  \times SU\left(  3\right)  $ becomes
\begin{equation}
\left(  1,1,1\right)  +\left(  1^{\prime},1,1\right)  +\left(  1,1,3\right)
+\left(  1^{\prime},1,3\right)  +\left(  1,2,1\right)  +\left(  1,2,3\right)
\end{equation}
These correspond to the particles, respectively, $\nu_{R}$, $e_{R},$
\ $u_{R},$ $d_{R},$ $l_{L},$ $q_{L}$ where $l_{L}=\left(
\begin{array}
[c]{c}%
\nu\\
e
\end{array}
\right)  _{L}$ is the lepton doublet and $q_{L}=\left(
\begin{array}
[c]{c}%
u\\
d
\end{array}
\right)  _{L}$ is the quark doublet.

\section{Spectral Action}

The Euclidean fermionic action, including all vertex interactions is extremely
simple and is given by
\begin{equation}
\left(  J\Psi,D_{A}\Psi\right)
\end{equation}
where the path integral is equal to the Pfaffian of the operator $D_{A},$
eliminating half of the degrees of freedom associated with mirror fermions
\cite{saddam}. In the Lorentzian form $J\Psi=\Psi$ and half of the degrees of
freedom are eliminated by the reality condition, showing the equivalence of
both cases when the $KO$ dimension of the space is $10$ \cite{Co}
\cite{Barrett}$.$ The bosonic action which gives the dynamics of all the
bosonic fields, including graviton, gauge fields and Higgs fields, is governed
by the spectral action principle which states that the action depends only on
the spectrum of the Dirac operator $D_{A}$ given by its eigenvalues which are
geometric invariants. The spectral action is given by%
\begin{equation}
\mathrm{Tr}\left[  f\left(  \frac{D_{A}}{\Lambda}\right)  \right]
\label{spectral}%
\end{equation}
where $f$ is a positive function and $\Lambda$ is a cut-off scale. At scales
below $\Lambda$ the function $f$ could be expanded in a Laurent series in
$D_{A}$ thus reducing evaluating the spectral action (\ref{spectral}) to that
of calculating the heat-kernel coefficients, which are geometric invariants.
Thus the spectral action, at energies lower than the cut-off scale is
determined by the Seeley-de Witt invariants of the operator $D_{A}$ with the
coefficients in the expansion related to the Mellin transform of the function
$f.$ The result for the Standard Model, at unification scale is given by
\begin{align*}
S_{\mathrm{b}}  &  =\frac{24}{\pi^{2}}f_{4}\Lambda^{4}%
{\displaystyle\int}
d^{4}x\sqrt{g}\\
&  -\frac{2}{\pi^{2}}f_{2}\Lambda^{2}%
{\displaystyle\int}
d^{4}x\sqrt{g}\left(  R+\frac{1}{2}a\overline{H}H+\frac{1}{4}c\sigma
^{2}\right) \\
&  +\frac{1}{2\pi^{2}}f_{0}%
{\displaystyle\int}
d^{4}x\sqrt{g}\left[  \frac{1}{30}\left(  -18C_{\mu\nu\rho\sigma}%
^{2}+11R^{\ast}R^{\ast}\right)  +\frac{5}{3}g_{1}^{2}B_{\mu\nu}^{2}+g_{2}%
^{2}\left(  W_{\mu\nu}^{\alpha}\right)  ^{2}+g_{3}^{2}\left(  V_{\mu\nu}%
^{m}\right)  ^{2}\right. \\
&  \qquad\left.  +\frac{1}{6}aR\overline{H}H+b\left(  \overline{H}H\right)
^{2}+a\left\vert \nabla_{\mu}H_{a}\right\vert ^{2}+2e\overline{H}H\,\sigma
^{2}+\frac{1}{2}d\,\sigma^{4}+\frac{1}{12}cR\sigma^{2}+\frac{1}{2}c\left(
\partial_{\mu}\sigma\right)  ^{2}\right]
\end{align*}
where $C_{\mu\nu\rho\sigma}$ is the conformal tensor, $B_{\mu},$ $W_{\mu
}^{\alpha},$ $V_{\mu}^{m}$ are the gauge fields of $U\left(  1\right)
_{Y}\times SU\left(  2\right)  \times SU\left(  3\right)  $, $H$ is a Higgs
doublet and $\sigma$ is a singlet. The coefficients $a,$ $b,$ $c,$ $d,$ $e$
are given in terms of the Yukawa couplings of the Higgs fields to the
fermions. A similar calculation for the unbroken algebra (\ref{PS}) will give
the bosonic action of Pati-Salam models including all the gauge and Higgs
interactions with their potential.

\section{Conclusions}

It is remarkable that the simple two sided relation (\ref{quanta}) leads to
volume quantization of the four-dimensional Riemannian manifold with Euclidean
signature. The manifold could be reconstructed as a composition of the
pullback maps from two separate four spheres with coordinates defined over two
Clifford algebras. The phase space of coordinates and Dirac operator defines a
noncommutative space of $KO$ dimension $10.$ The symmetries of the algebras
defining the noncommutative space turn out to be those of $SU\left(  2\right)
_{R}\times SU\left(  2\right)  _{L}\times SU\left(  4\right)  $ known as the
Pati-Salam models. Connections along discrete directions are the Higgs fields.
A special case of this configuration occurs when the order one condition
(\ref{order}) is satisfied, reducing the finite algebra to the subalgebra
given by (\ref{SM}). The action has a very simple form given by a Dirac action
for fermions and a spectral action for bosons. The $16$ fermions (per family)
are in the correct representations with respect to Pati-Salam symmetries or
the SM symmetries. There are many consequences of the volume quantization
condition which could be investigated. For example imposing the quantization
condition through a Lagrange multiplier would imply that the cosmological
constant will arise as an integrating constant in the equations of motion. One
can also look at the possibility that only the three volume (space-like) is
quantized. This can be achieved provided that the four-dimensional manifold
arise due to the motion of three dimensional hypersurfaces, which is
equivalent to the $3+1$ splitting of a four-dimensional Lorentzian manifold.
Then three dimensional space volume will be quantized, provided that the field
$X$ that maps the real line have a gradient of unit norm $g^{\mu\nu}%
\partial_{\mu}X\partial_{\nu}X=1.$ It is known that this condition when
satisfied gives a modified version of Einstein gravity with integrating
functions giving rise to mimetic dark matter \cite{CM} \cite{CMV}. All of this
could be considered as a first step towards quantizing gravity. What is done
here is the analogue of phase space quantization where the Dirac operator
plays the role of momenta and the maps $Z$ play the role of coordinates. The
next step would be to study consequences of this new idea on the quantization
of fields which are functions of either coordinates $Z$ or Dirac operator
$D_{A}.$ All of this and other ideas will be presented elsewhere.

To conclude, the influence of Salam on my work which started with
supersymmetry, supergravity and their applications continued to string theory,
topological gravity and noncommutative geometry, cannot be underestimated. It
is my honor and privilege to have followed his footsteps. It was my good
fortune to have him as my thesis advisor at Imperial College and postdoctoral
mentor at ICTP.

\section*{Acknowledgement}

I would like to thank Alain Connes for enjoyable collaboration during the last
twenty years. I\ would also like to thank Slava Mukhanov for his friendship
and collaboration. The joint work with Walter van Suijlekom opened a new
avenue in developing our program. This work is supported in part by the
National Science Foundation under Grant No. Phys-1202671 and Phys-1518371.

\end{document}